%% file: Lattice2012_proceedings.tex
\documentclass{PoS}
\usepackage{amsmath}
\usepackage{slashed}
\usepackage[caption=false]{subfig}	

\title{Higgs boson mass bounds in the presence of a heavy fourth quark family}

\ShortTitle{Higgs boson mass bounds in the presence of a heavy fourth quark family}

\author{John Bulava\\
        CERN Theory Devision\\
        E-mail: \email{john.bulava@cern.ch}}

\author{Philipp Gerhold, Jim Kallarackal, \speaker{Attila Nagy}\\
        Humboldt University Berlin\\
        E-mail: \email{philipp\_gerhold@hotmail.com}, \\ \email{Jim.kallarackal@physik.hu-berlin.de},\\ \email{nagy@physik.hu-berlin.de}}

\author{Karl Jansen\\
        NIC, Desy Zeuthen \\
        E-mail: \email{karl.jansen@desy.de}}

\abstract{We present Higgs boson mass bounds in a lattice regularization allowing 
thus for non-perturbative investigations. In particular, we employ a lattice modified 
chiral invariant Higgs-Yukawa model using the overlap operator. 
We show results for the upper and lower Higgs boson mass bounds in the presence of 
a heavy mass-degenerate quark doublet with masses ranging up to 700 GeV. 
We perform infinite volume extrapolations in most cases, and examine several 
values of the lattice cutoff. Furthermore, we argue that the lower Higgs boson mass
bound is stable with respect to the addition of higher dimensional 
operators to the scalar field potential. Our results have severe consequences 
for the phenomenology of a fourth generation of quarks if a light Higgs boson is discovered at the LHC.
			}

\FullConference{The 30th International Symposium on Lattice Field Theory\\
                 June 24 -- 29,  2012\\
                 Cairns, Australia}

\begin{document}

\section{Introduction}
\input{Lattice2012_introduction.tex}

\section{Model and Implementation}
\input{Lattice2012_model_and_implementation.tex}
\section{Results}
\input{Lattice2012_results.tex}
\section{Conclusion and Outlook}
\input{Lattice2012_conclusion.tex}

\bibliographystyle{unsrt}
\bibliography{Lattice2012_refs.bib}

\end{document}

%% file: Lattice2012_introduction.tex
The existence 
of the standard model Higgs boson  
is a necessary ingredient for the consistency of the standard model. 
Although the mass of the Higgs boson cannot be predicted,  
bounds on the Higgs boson mass can be given. 
A lower bound can be derived from the requirement 
of a stable vacuum while an upper bound can be computed from 
the triviality of the theory which leads to the occurrence of the 
Landau pole. 
These mass bounds are usually computed in perturbation theory and 
indeed, perturbative results for the Higgs boson mass bounds were obtained in the past. 
However, the perturbative analysis may be questioned, since for the 
upper bound, the quartic coupling can become strong. The lower 
bound, on the other hand, may be an artefact of perturbation theory
when looking at Higgs field values far away from the minimum of 
the potential.  
These concerns motivated a non-perturbative and ab initio 
lattice field theory investigation of 
the mass bounds of the Higgs boson in a chirally invariant 
lattice Higgs-Yukawa model \cite{Gerhold:2009ub,Gerhold:2010bh}.

In the lattice approach it is straightforward to determine 
those mass bounds also for the case that there is a fourth, 
heavy family of quarks. Such an extension is not excluded a 
priori and further, a fourth family of fermions offers the possibility 
of generating sufficient CP violation to fulfil
the Sakharov condition to 
explain electroweak baryogenesis of the early universe \cite{Holdom:2009rf}.

However, CKM4 fits and direct searches already make a naive fourth 
generation scenario quite improbable \cite{Eberhardt:2012gv}, 
especially if the newly discovered particle with a mass around 
126~GeV \cite{:2012gu,:2012gk} turns out to be the standard model 
Higgs boson. In this work we provide additional general 
constraints on the fourth family quark masses by their 
effect on the Higgs boson mass bounds.

%% file: Lattice2012_model_and_implementation.tex
We study the Higgs-Yukawa sector of the standard model. The field content we consider is a fermion doublets $\Psi$ and a complex scalar doublet $\varphi$. The continuum action in this model is given by:
\begin{multline}\label{eq:action_continuum}
	S^{\text{cont}}[\bar{\psi}, \psi, \varphi] = \int d^4 x \left\{\frac{1}{2}\left(\partial_{\mu} \varphi \right)^{\dagger} 
	\left(\partial^{\mu} \varphi \right)
	+  \frac{1}{2} m_0^2 \varphi^{\dagger} \varphi + \lambda_0 \left(\varphi^{\dagger} \varphi \right)^2 \right\} \\
			+\int d^4 x  \left\{\bar{t} \slashed \partial t + \bar{b} \slashed \partial b +
			y_{b_0}\bar{\psi}_L \varphi\, {b}_{_R} +
			y_{t_0}\bar{\psi}_L \tilde \varphi\, {t}_{_R}
			+ h.c. \right\},\quad \text{with }\tilde \varphi=i\tau_2 \varphi^*.
\end{multline}
The bare quartic self coupling of the scalar field is given by $\lambda_0$, $m_0$ denotes the 
bare mass of the scalar field and $y_{t_0/b_0}$ are the Yukawa couplings of the 
fermion fields. We want to stress, that we do not include any gauge fields in this 
model neither the gluonic degrees of freedom nor electroweak ones. 
This is done for computational simplicity and further we expect the effect of 
the gauge fields to be small for the problems addressed in our work.

We will mostly study the system in the phase with spontaneous symmetry breaking
where the scalar field develops a non vanishing vacuum expectation value ($vev$)
and three Goldstone modes emerge. Further we only consider the heaviest fermion doublet, 
since the dynamics of the scalar field is dominated by the largest Yukawa coupling.

For the discretized lattice version of the bosonic action we rewrite the scalar 
doublet as a real four vector $\Phi$. With the reparametrisations
\begin{equation}\label{eq:reparametrisation_bosonic}
	 a \varphi = \sqrt{2 \kappa} \left( \begin{array}{c} \Phi^2 + i\Phi^1 \\ \Phi^0 - i \Phi^3 \end{array} \right) ,\quad\quad
	 \lambda_0 = \frac{\hat{\lambda}}{{4 \kappa^2}},\quad\quad
	 a^2 m_0^2 = \frac{1 - 2 \hat{\lambda} -8 \kappa}{\kappa}
\end{equation}
the bosonic action can then be written in the compact form:
\begin{equation}\label{eq:action_lattice_bosonic}
	S_B[\Phi] = -\kappa \sum\limits_{x,\mu} \Phi_x^{\dagger} \left[\Phi_{x+\mu} + \Phi_{x-\mu}\right] + 
				\sum\limits_{x} \Phi_x^{\dagger} \Phi_x + 
				\hat{\lambda}\sum\limits_{x} \left[ \Phi_x^{\dagger} \Phi_x - 1 \right]^2.
\end{equation}
Here $\kappa$ denotes the hopping parameter. The subscripts $x$ and $x\pm\mu$ on the scalar 
fields denote the scalar field at the space-time point $x$ and $x\pm a \hat\mu$ respectively.

For the discretization of the fermions we use the overlap 
operator $D^{ov}$ \cite{Neuberger:1997fp, Neuberger:1998wv, Hernandez:1998et} with a Wilson Dirac kernel $D^W$:
\begin{equation} \label{eq:DefOfNeuberDiracOp}
 D^{ov}  =  \rho \left\{1+\frac{ A}{\sqrt{ A^\dagger  A}}   \right\},  \quad\quad A = D^{W} - \rho,\quad\quad
 D^{W} = \sum\limits_\mu \gamma_\mu \nabla^s_\mu - \frac{r}{2} \nabla^b_\mu\nabla^f_\mu,
\end{equation}
with $\nabla^{f,b,s}_\mu$ being the forward, 
backward and symmetrized lattice nearest neighbor difference operators in 
direction $\mu$. The so-called Wilson parameter $r$ is as usual chosen to be $r=1$. 
The dimensionless parameter $\rho$ is free to be chosen in the range $0 < \rho < 2r$. 
However, the locality properties of the free overlap operator in 
case of vanishing gauge couplings are optimal for the case of 
$\rho=1$ \cite{Hernandez:1998et}, which will therefore be the choice in our work.
The fermionic part of the action is then:
\begin{equation}\label{eq:action_lattice_fermionic}
	S_F[\bar{\psi},\psi,\Phi] = \sum\limits_{x} \bar{\psi}_x\left[ D^{ov} + P_{_+} \phi_x^{\dagger} \text{diag}(\hat{y}_t,\hat{y}_b) 
	\hat{P}_{_+} + 
						P_{_-} \text{diag}(\hat{y}_t,\hat{y}_b) \phi_x \hat{P}_{_-} \right] \psi_x, \quad \hat y_{t/b} = \sqrt{2 \kappa}\, y_{{t/b}_0}.
\end{equation}
This action now obeys an exact global $\mbox{SU}(2)_L\times \mbox{U}(1)_Y$ 
lattice chiral symmetry with the transformations:
\begin{equation}
 \psi \rightarrow  U_Y \hat P_+ \psi + U_Y\Omega_L \hat P_- \psi ,\quad
 \bar\psi \rightarrow \bar\psi P_+ \Omega_L^\dagger U_{Y}^\dagger + \bar\psi P_- U^\dagger_{Y}, \quad
\phi \rightarrow U_Y  \phi \Omega_L^\dagger, \quad
\phi^\dagger \rightarrow \Omega_L \phi^\dagger U_Y^\dagger,
\end{equation}
for any $\Omega_L \in \text{SU}(2)_L$ and $U_Y \in \text{U}(1)_Y$.
The modified chiral projectors are given by:
\begin{equation}
 \hat{P}_{_{+/-}} = \frac {1 \pm \hat{\gamma}^5}{2}, \quad \hat{\gamma}^5 = \gamma^5 \left( 1 - \frac{1}{\rho} D^{ov} \right)
\end{equation}

Even though in principle different masses for the fermions in the doublet are 
possible we restrict ourselves to a mass-degenerate doublet in this work.
For the implementation we use a polynomial Hybrid Monte Carlo 
algorithm \cite{Frezzotti:1997ym} with various improvements implemented. 
For further details of the implementation see \cite{Gerhold:2010wy}.

To set the scale, i.e.\ to determine the lattice spacing $a$, we use the phenomenologically 
known value of the $vev$ of 246~GeV. Further we define the cutoff $\Lambda$ as the 
inverse of the lattice spacing. Since in finite volume a naive average scalar field 
would vanish without an external field, we define the magnetization $m$ as the 
average absolute value of the scalar field:
\begin{equation}
 m = \left< | \bar \Phi | \right>,\quad\quad \bar \Phi = V^{-1} \sum_x\Phi_x,
\end{equation}
with $V$ being the volume of the space time lattice. This approach has the same 
thermodynamical limit as the application of an external source \cite{Gockeler:1991ty}. 
The magnetization and the renormalized $vev$ are related as follows:
\begin{equation}\label{eq:scale_setting}
 v_r = \frac{m \sqrt{2 \kappa}}{\sqrt{Z_G}},\quad\quad    \frac{v_r}{a} = 246 \text{ GeV},\quad\quad \Lambda=\frac{246 \text{ GeV}}{v_r}.
\end{equation}
%
%
The Goldstone and Higgs bosom renormalization constants $Z_{G/H}$ and their masses are computed 
from the real part of the Goldstone or Higgs boson propagators $G_{G/H}$ respectively:
\begin{equation}
 Z_{G}^{-1} =  \left. \frac{d}{d(p^2)} \Re \left( G_{G}^{-1}(p^2) \right) \right|_{p^2=-m^2_{G}}, \quad\quad 
 \left.\Re\left(G_{G/H}^{-1}(p^2)\right)\right|_{p^2=-m^2_{H/G}} = 0.
\end{equation}
The propagators are computed for discrete lattice momenta and fitted according 
a one loop motivated formula derived in lattice perturbation theory. It may be noted, 
that this approach especially for the Higgs boson is only valid if the decay width of 
the particle is small compared to its mass. However, it was shown in a rigorous 
resonance analysis \cite{Gerhold:2011mx}, that at least in the case of a physical 
top quark mass, the pole of the propagator fully agrees with the resonance mass 
determined in \cite{Gerhold:2011mx}. 
Finally, the masses of the fermions $m_f$ are computed by means of the time slice correlator
\begin{equation}
 C_f(\Delta t) = \frac{1}{L_t \cdot L_s^6} \sum_{t=0}^{L_t-1}\sum_{\vec{x}, \vec{y}} 
                \left \langle 2 \Re \operatorname{Tr} \left \{ 
					 \hat{P}_{_L}\psi (t+ \Delta t, \vec x) \cdot \bar{\psi}(t, \vec y)  P_{_L} 
					  \right\} \right \rangle,
\end{equation}
which shows a behavior proportional to $\cosh( a m_f ( \Delta t - L_t/2) )$ for large time separations $\Delta t$.

\begin{figure}
\centering
\subfloat[magnetization]{\includegraphics[width=0.33\linewidth]
{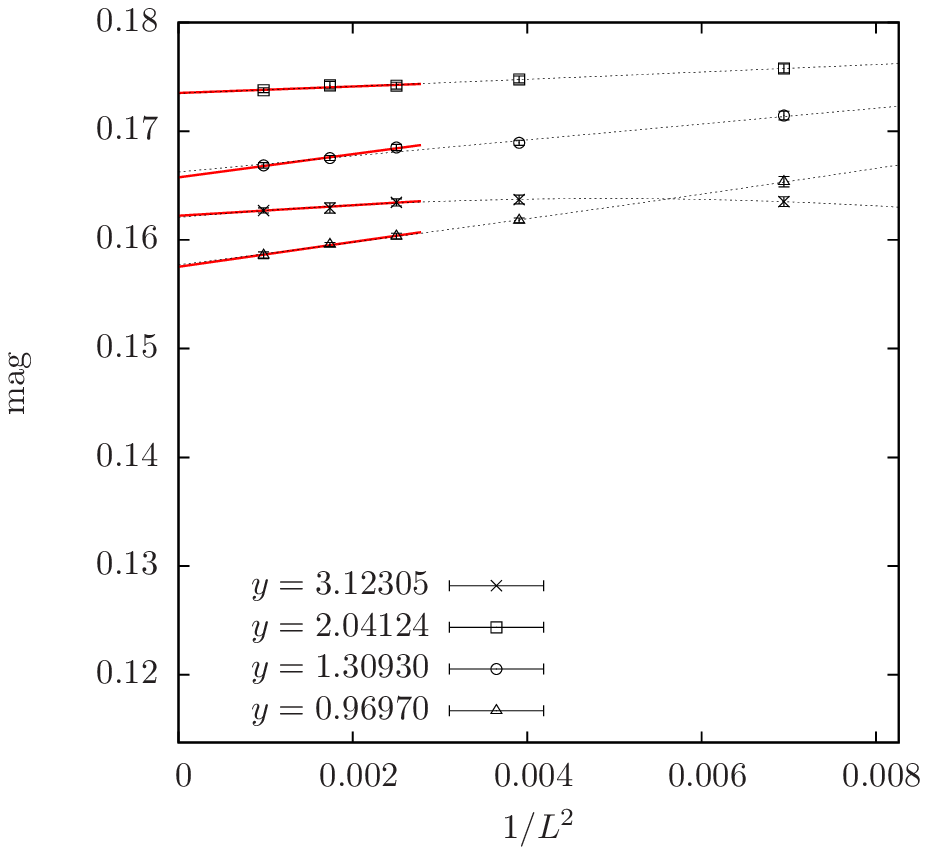}} 
\subfloat[top quark mass]{\includegraphics[width=0.33\linewidth]
{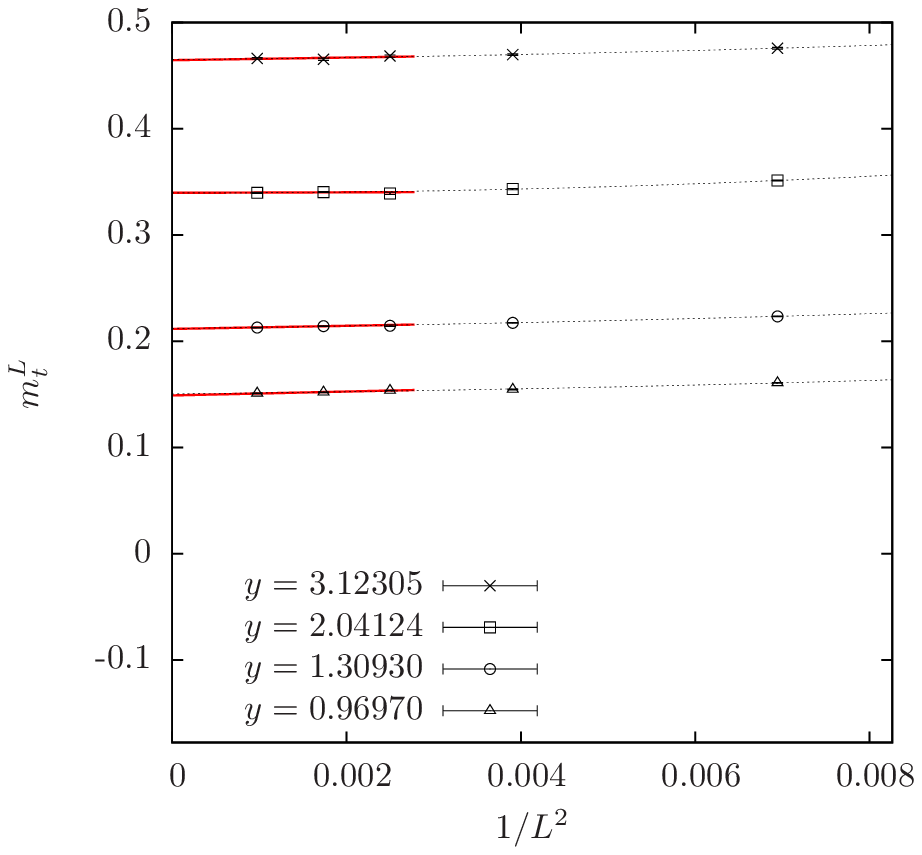}}
\subfloat[Higgs boson mass]{\includegraphics[width=0.33\linewidth]
{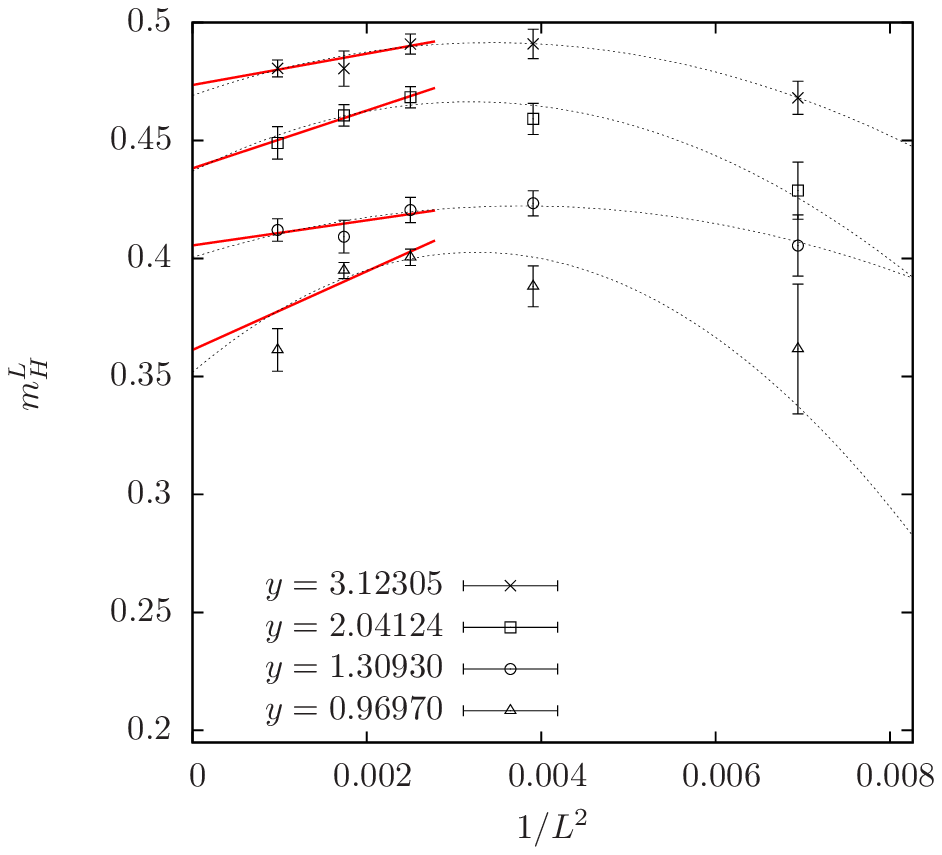}}
\caption{We show finite volume effects for the magnetization, the quark mass and the Higgs boson mass 
at a cutoff of about 1.5 TeV. The data shown correspond to simulations 
performed at infinite bare quartic coupling and fermion masses in the range $m_{t'} \approx 200 \dots 700$ GeV. 
The lattice sizes shown 
are $L_s=12,16,20,24,32$. The red lines indicate linear fits involving only data 
from computations with $L_s \geq 20$ the grey lines show quadratic fits over the whole range of lattice extents.}
\label{fig:finiteVolumeExamples}
\end{figure}

In general we observe rather severe finite volume effects due to the almost 
massless Goldstone modes, which cause finite volume effects 
proportional to $L_s^{-2}$ \cite{Hasenfratz:1989pk, Hasenfratz:1990fu} rather 
than an exponential falloff with Euclidean time for theories with a mass gap. 
Some examples for finite volume effects can be found in fig.~\ref{fig:finiteVolumeExamples}. 
There, one can see, that for a trustworthy determination of the Higgs boson mass 
and the $vev$
lattice sizes of up to at least $L_s=32$ are necessary.

In addition to the non-perturbative determination we also perform a perturbative analysis 
by means of the constrained effective potential as described in detail in \cite{Gerhold:2010wy}. 
Those calculations are performed within the same lattice regularization as the numerical simulations. 
We employ discrete 
lattice sums for the loop corrections which are then evaluated numerically. 
The effective potential $V(\bar \phi)$ is determined to one loop in the large $N_f$ limit 
with the $vev$ and the Higgs boson mass given by:
\begin{equation}
 \left. \frac{d\, V(\bar \phi)}{d\, \bar \phi}\right|_{\bar\phi=v} =0; \quad \quad
 \left. \frac{d^2\, V(\bar \phi)}{d\, \bar \phi ^2}\right|_{\bar\phi=v} =m_H^2.
\end{equation}
In order to have a stable vacuum in the scaling regime ($|\bar\phi|<0.5$), we demand the 
potential to be concave up everywhere in this region. In this perturbative 
framework we can determine the lower Higgs boson mass bound by finding the lowest 
possible Higgs boson mass while keeping physical quantities fixed and 
still fulfil the required stability conditions.

%% file: Lattice2012_results.tex
\begin{figure}
\centering
\subfloat[physical top quark mass]{\includegraphics[width=0.38\linewidth]
{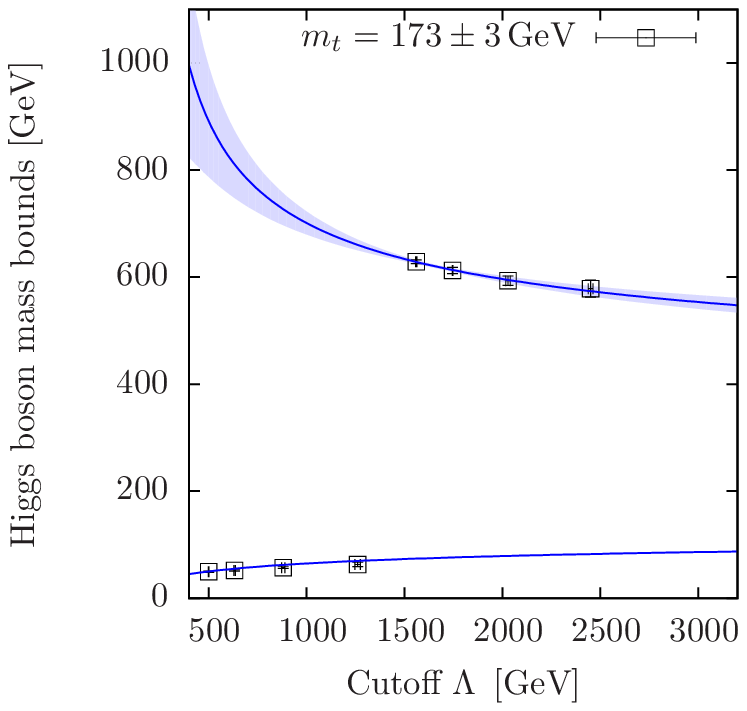}} \hspace{0.14\linewidth} 
\subfloat[heavy quark double]{\includegraphics[width=0.38\linewidth]
{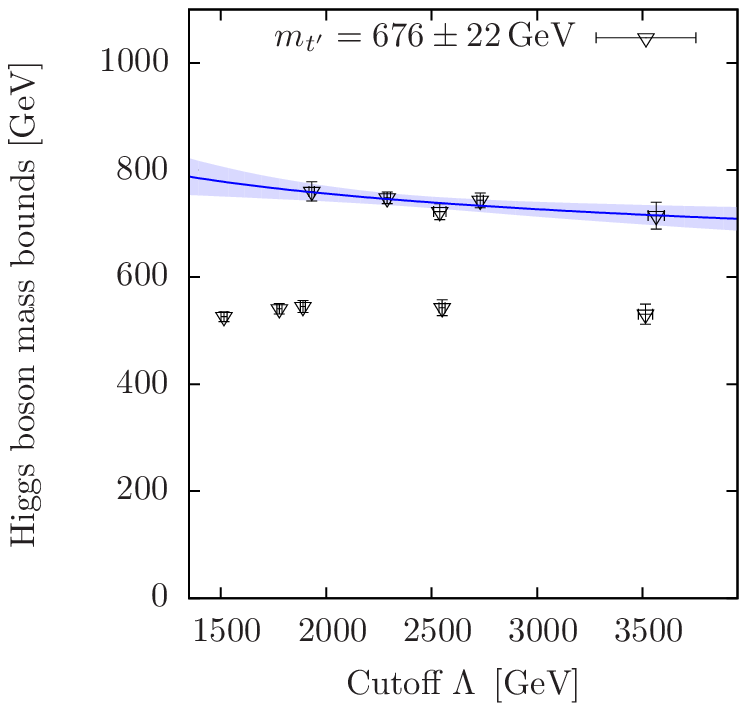}}
\caption{We show the cutoff dependence of the Higgs boson mass bounds for two masses of a mass degenerate quark doublet. 
On the left graph we show the data for the physical top quark mass while the right graph corresponds to a 
very heavy mass around 676 GeV. }
\label{fig:cutoffDependence}
\end{figure}
\begin{figure}
\centering
\subfloat[Results at $\Lambda\approx 1.5$ TeV]{\label{fig:massbounds}\includegraphics[width=0.49\linewidth]
{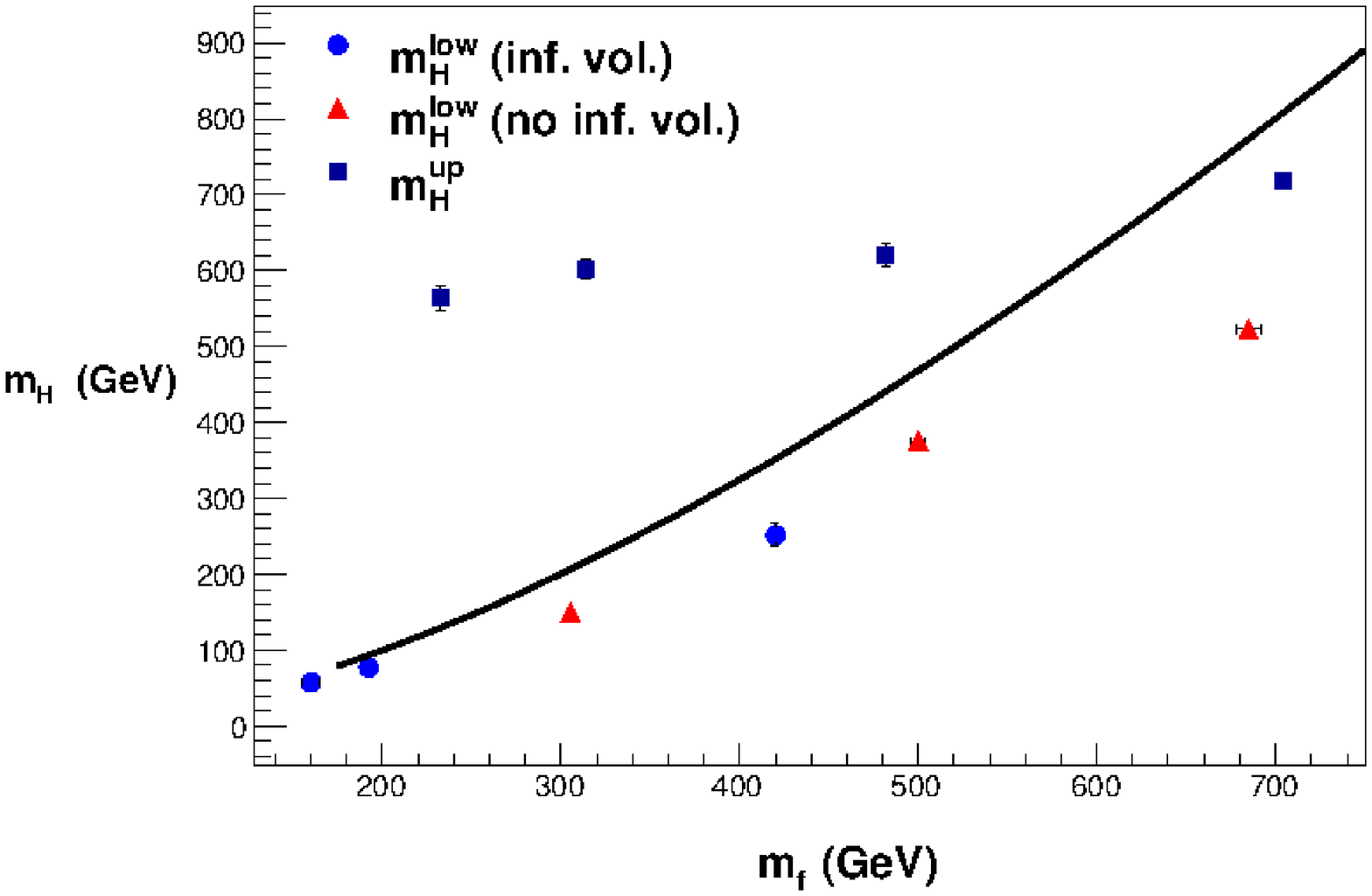}} 
\subfloat[Higher dimensional operators]{\label{fig:higherDimOps}\includegraphics[width=0.49\linewidth]
{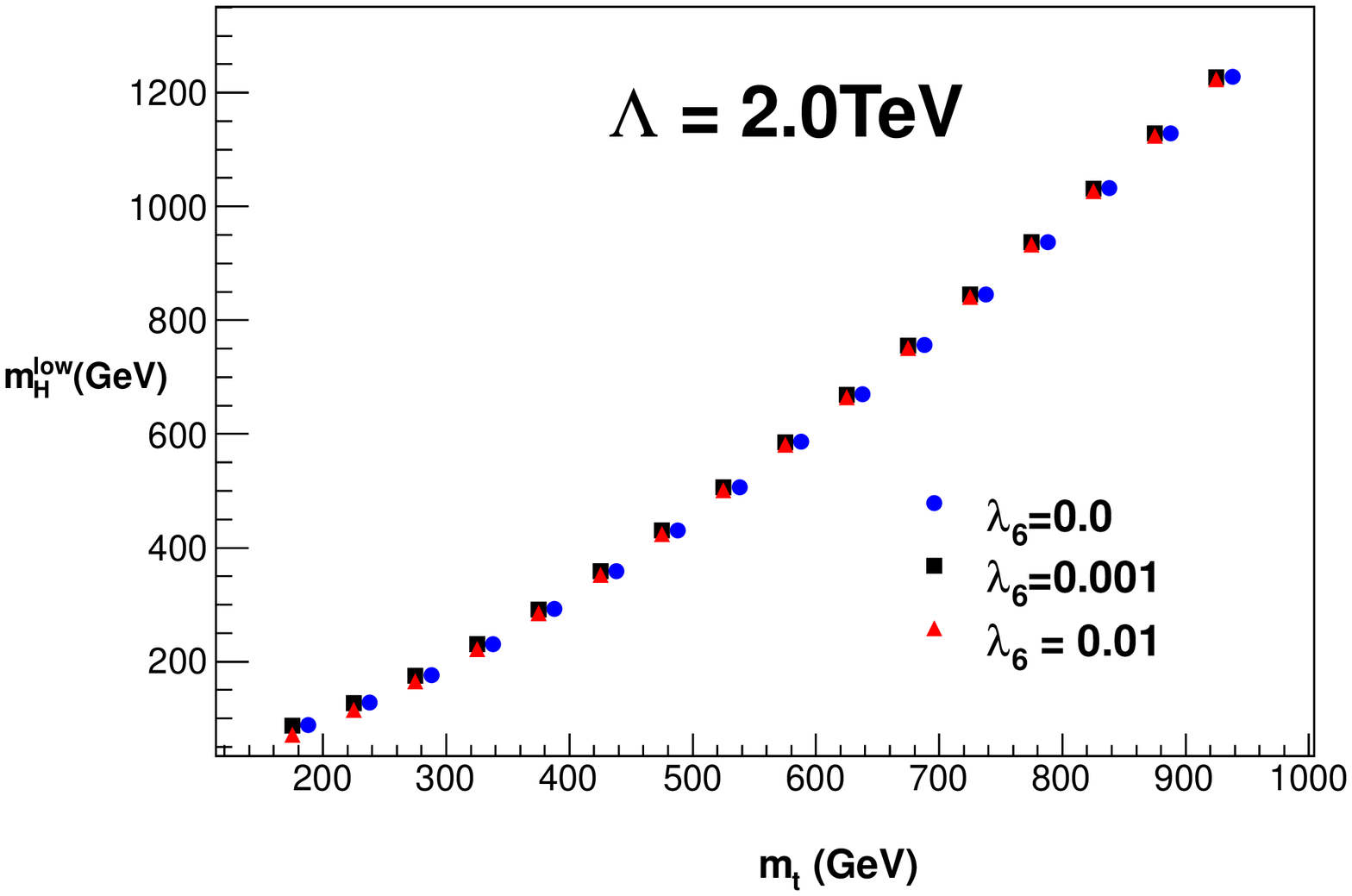}}
\caption{We show Higgs boson mass bounds at fixed cutoff. On the left graph non-perturbative data 
are shown for the upper and lower mass bound. 
(For the red triangles the infinite volume extrapolation has not been carried out yet). 
The data points for the lower bound are compared to findings in a perturbative effective potential calculation. 
The right-hand plot compares perturbative results for the lower bound with and without a 
dimension-6 operator included at a cutoff $\Lambda=2$ TeV. }
\end{figure}

Let us shortly summarize the strategy to determine mass bounds for the 
Higgs boson in our analysis. In total we have three free parameters, 
namely the bare quartic self coupling $\hat\lambda$, the hopping parameter 
$\kappa$ and the Yukawa coupling $\hat y$. We fix the Yukawa coupling such to 
obtain the desired mass for the fermions and tune $\kappa$ to retrieve the cutoff 
at which we want to investigate the mass bounds. Then it can be shown  
\cite{Gerhold:2009ub,Gerhold:2010bh}, 
that with the cutoff and the fermion mass fixed, the smallest accessible 
Higgs boson mass is obtained for choosing $\hat \lambda =0$ while the 
Higgs boson mass is the largest for $\hat \lambda =\infty$. In earlier 
works we investigated the cutoff dependence of the Higgs boson mass 
bounds for the case of a fermion doublet at the physical top quark mass 
$m_t \approx 175$ GeV \cite{Gerhold:2009ub, Gerhold:2010bh} and for a 
very heavy doublet with a mass around $m_{t'} \approx 676$ GeV \cite{Gerhold:2010wv}. 
The results can be found in fig.~\ref{fig:cutoffDependence}. While the upper 
Higgs boson mass bound is only increased slightly, the lower mass bound for the 
Higgs boson increased by a factor around 5 when the fermion mass is increased. 
To obtain a better understanding of the increase in the mass bounds, we 
investigated those bounds for several fermion masses at a cutoff around 
$\Lambda \approx 1.5 \text{ TeV}$. Those results are shown in fig.~\ref{fig:massbounds}. 
One clearly observes the smooth increase of the lower mass bound of the 
Higgs boson with increasing quark mass. Our results suggest that a Higgs boson 
mass of 
126~GeV would allow heavy quarks only up to approximately 300~GeV.

In fig.~\ref{fig:massbounds} we also show 
the lower Higgs boson mass bound obtained from the 
perturbative effective potential calculations. It is remarkable, how well the 
perturbative result agrees qualitatively with the non-perturbative findings up to 
quark masses of about 700~GeV (corresponding to a Yukawa coupling around 3). 
This lets us assume, that the model may be perturbative in a wide region of the 
parameter space. This finding allows us to test perturbatively whether adding a
dimension-6 operator, i.e. a $\lambda_6 \varphi^6$ term, to the bosonic action 
may have an impact on the lower bound. To this end, we determined the lower bound 
at a cutoff $\Lambda = 2\text{~TeV}$ for various quark masses with and without a 
dimension-6 operator at two different couplings $\lambda_6$. Those results are shown in 
fig.~\ref{fig:higherDimOps}. We clearly see that the additional operator has no 
visible impact on the lower Higgs boson mass bound.

%

%% file: Lattice2012_conclusion.tex
We performed a non-perturbative determination of Higgs boson mass bounds in a 
Higgs-Yukawa model with a physical top quark mass and a heavy quark doublet. 
We found that a 126~GeV Standard Model Higgs boson excludes a naive fourth generation of 
quarks if their mass exceeds 300 GeV. Further we gave perturbative arguments, 
that this bound is stable against the inclusion of higher dimensional operators. 
However, we plan to addressed this finding also non-perturbatively in the future. 
Further, our setup allows to test whether a non-degenerate doublet could alter the mass bounds, 
since a mass splitting of up to $m_{b'}/m_{t'} \approx 0.9$ is possible \cite{Denner:2011vt}. 
Another interesting direction would be to include 
an additional scalar field, as suggested in \cite{EliasMiro:2012ay}, 
which could have an impact on the Higgs boson mass bound.

\section*{Acknowledgements}
This work was supported by the DFG through the DFG-project Mu932/4-4. The numerical Simulation have been performed at the 
SGI system HLRN-II at the HLRN supercomputing service in Berlin and Hannover and on the HPC cluster at DESY Zeuthen.